\documentclass[review]{elsarticle}
\usepackage{bm}
\usepackage{amssymb}
\usepackage{times}
\usepackage{latexsym}
\usepackage{graphicx}
\usepackage{color}
\usepackage{subfigure}
\usepackage{amsmath}
\journal{Phys. Lett. A}

\usepackage{times}
\begin{document}

\title{Topological Map of the Hofstadter Butterfly and\\
 Van Hove Singularities }

\author{Gerardo Naumis$^{1}$,$^{2}$} 
\ead{naumis@fisica.unam.mx}
\author{Indubala I Satija$^{2}$}

\address{1. Departamento de F\'{i}sica-Qu\'{i}mica, Instituto de
F\'{i}sica, Universidad Nacional Aut\'{o}noma de M\'{e}xico (UNAM),
Apartado Postal 20-364, 01000 M\'{e}xico, Distrito Federal,
M\'{e}xico}
\address{2. Department of Physics  and Astronomy , George Mason University, Fairfax, Virginia 22030, USA }

\begin{abstract}
The Hofstadter butterfly is a quantum fractal with a highly complex nested set of gaps, where each gap represents a quantum Hall state whose quantized conductivity is characterized by topological invariants known as the Chern numbers.
Here we obtain simple rules to determine the Chern numbers  at all scales  in the butterfly fractal and lay out a very detailed topological map of the butterfly.
Our study reveals the existence of a set of critical points,  each corresponding to a macroscopic annihilation of  orderly patterns of both the positive and  the negative Cherns that appears as a fine structure in the butterfly. 
Such topological collapses are identified with theVan Hove singularities that exists at every band center in the butterfly landscape.  We thus associate a topological character to the Van Hove anomalies. Finally, we show that this fine structure is amplified
 under perturbation,  inducing quantum phase transitions  to higher Chern states in the system.
\end{abstract}
\maketitle

Discovered by Belgian physicist Leon Van Hove in 1953,  Van Hove singularities are  singularities in the density of states  of a crystalline solid\cite{VH}.  These singularities are known to be responsible
for various anomalies provided Fermi level lies close to such a singularity.
Electronic instabilities at the crossing of the Fermi energy with a Van Hove singularity in the density of states ($DOS$) often lead to new phases of matter such as superconductivity,
magnetism,  or density waves\cite{VHS}.\\

A two-dimensional electron gas (2DEG) in a square lattice provides a simple example of Van Hove singularities in the energy dispersion of a  crystal.
For a tight binding model of a square lattice
, the energy dispersion is given by,
\begin{equation}
E = -2J [ \cos k_x a + \cos k_y a ]
\end{equation}

\begin{figure*}
\includegraphics[width =.6\linewidth,height=0.5\linewidth]{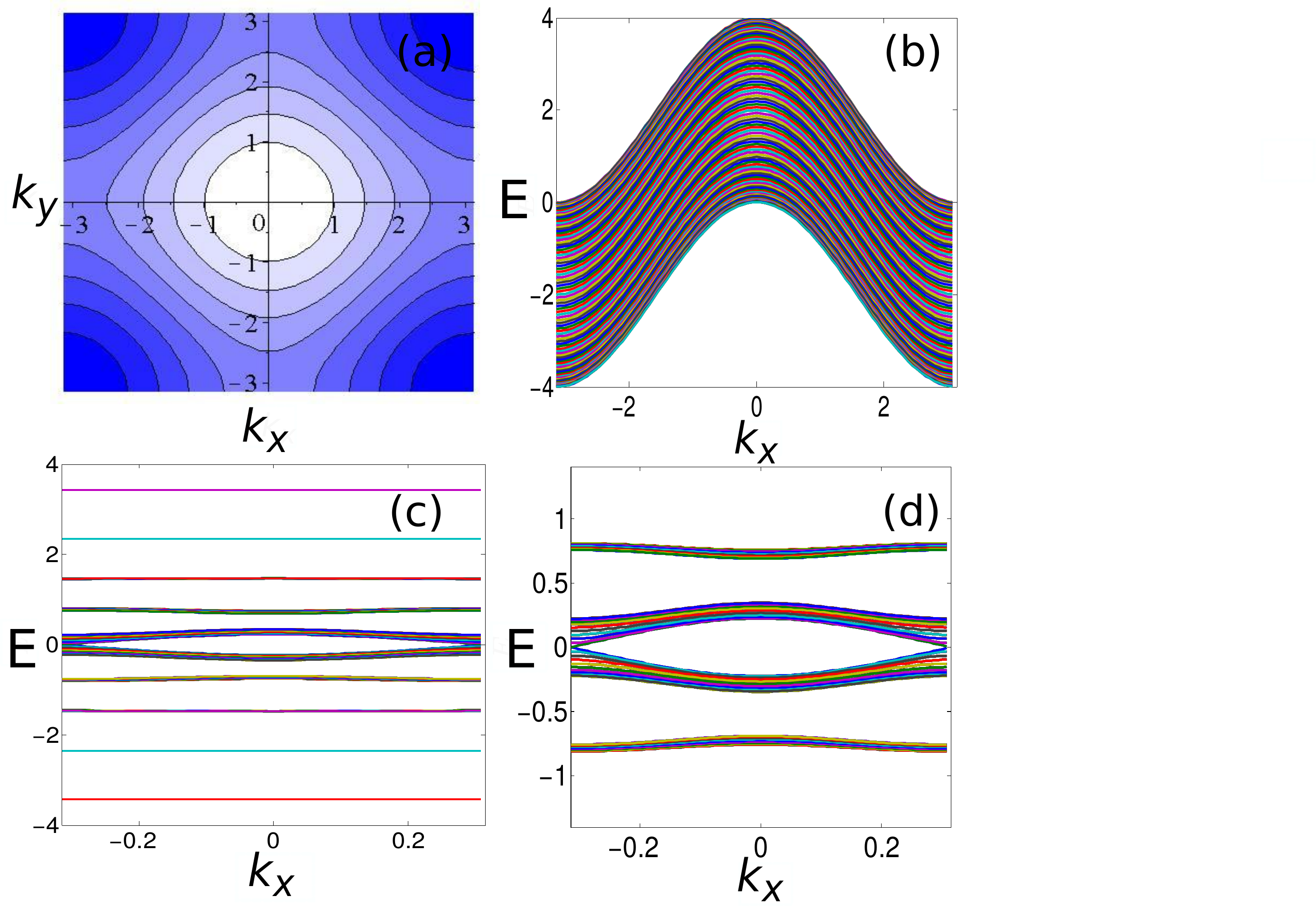}
\leavevmode \caption {(a) Contour plot of the energy $E$ in the $(k_x,k_y)$ plane, 
illustrating the saddle character of the band center for a 2DEG on a square lattice. (b) Shows the corresponding band as a function of $k_x$.  (c) shows the spectrum for small magnetic flux $\phi=0.1$.  Landau
levels correspond to the horizonal flat bands. In (d)  we show a blow up of c) near the band center, illustrating the deviation from  the Landau level picture near the band center that hosts a Van Hove singularity.  In (b),(c) and (d), the colours represent different values of $k_y$.  }
\label{EV}
\end{figure*}

Here $\vec{k}=(k_x, k_y)$ is the wave vector in the first Brillouin zone and $a$ is the lattice spacing of the square lattice and $J$ is the nearest-neighbor hopping parameter
which defines the effective mass  $m_e$ of the electron on the lattice by the relation $J=\frac{\hbar^2}{2m_e a^2}$. This single band Hamiltonian has band edges at $E=\pm 4J$.
It can be shown that the density of states (DOS)  at the band edges approaches a constant  equal to $\frac{1}{4 \pi a^2 \hbar^2}$. However, it diverges at the band center as
DOS $ \approx  \frac{ln J}{E}$. Such a divergence is an example of a Van Hove singularity. 
Figure \ref{EV} shows the energy contours in $(k_x,k_y)$ plane, where the almost free-electron concentric circles are transformed into a diamond shape structure that corresponds to saddle points in the energy surface.
We note that
the lattice structure is essential for the existence of Van Hove singularities.
Van Hove singularities have been given a topological interpretation in terms of a switching of electron orbits from electron like to hole like\cite{Markiewicz}.\\

In this paper we investigate the Van Hove anomalies of a 2DEG in transverse magnetic fields. Such system describes all phases of non-interacting electrons as one varies the chemical potential and magnetic field.
The phase diagram, known as the Hofstadter butterfly\cite{Hof} represents various quantum Hall states, each characterized by a quantum number, the Chern number, that has its roots in the nontrivial topology
of the underlying Hilbert space\cite{TKKN}.   The key result of this paper is the topological characterization of Van Hove singularities that are nested in the hierarchical pattern of the butterfly spectrum.
We show that in the two-dimensional energy-flux space, every vicinity of a Van Hove
 consists of interlacing sequences of positive and negative Chern numbers that collide and annihilate at the Van Hove singularities. In other words,  Van Hove singularities induce a topological collapse
 in the quantum fractal of  the Hofstadter butterfly. We calculate Chern numbers in the neighborhood of Van Hove singularities, facilitated by  simple rules that we derive
 for determining the entire topological map of the butterfly fractal at all scales.
 Our analysis begins with a geometrical approach, based on simple number theory, that sets the stage for determining the Chern numbers of all the gaps and its associated fine structure. It is the orderly patterns
 of  topological integers that characterize the fine structure that gets linked to the
Van Hove  anomalies of a two dimensional crystalline lattice in a magnetic field. In other words, we  correlate  the 
macroscopic nature of the topological collapse from which  emerges a reincarnation of Van Hove in a quantum fractal made up of integers
characterizing the quantum Hall conductivity.\\

A very recent study of the 2DEG when subjected to a  weak magnetic field\cite{kiv}, revealed the importance of Van Hove singularities in inducing changes in localization characteristics of the system. 
 In a continuum system, that is, in the absence of any lattice structure, the magnetic field  $B$ introduces a magnetic length $l_B = \sqrt{\frac{\phi_0}{ 2\pi B}}$
, reincarnation  of the cyclotron radius of the corresponding classical problem. In this limit, the energy spectrum consists of equally spaced harmonic oscillator levels known as the Landau-levels. Interestingly,
in a lattice with weak magnetic flux, the  Landau level picture breaks down near the band center as illustrated in the Fig. (\ref{EV}).
The insight into this clustering of the band fragments near the center can be gained by expanding the energy $E$ near $k_x+ k_y = \pm \pi$  and $k_x-k_y = \pm \pi$, where the energy depends linearly on the wave vector. This is in sharp
contrast to the quadratic dependence of energy near the band edges that leads to  simple harmonic levels,  namely the Landau levels.\\

 The model system that we study here consists of (spinless) fermions in a square lattice. 
Each site is labeled by a vector ${\bf r}=n\hat{x}+m\hat{y}$, where $n$, $m$ are
integers, $\hat{x}$ ($\hat{y}$) is the unit vector in the $x$ ($y$) direction, and $a$ is the lattice spacing. The tight binding Hamiltonian has the form
\begin{equation}
H=-J_x\sum_{\bf r}|\mathbf{r}+\hat{x} \rangle\langle \mathbf{r}|
-J_y\sum_{\bf r}|\mathbf{r}+\hat{y} \rangle e^{i2\pi n\phi} \langle \mathbf{r}|
+h.c. \label{qh}
\end{equation}
Here, $|\mathbf{r}\rangle$ is the Wannier state localized at site $\mathbf{r}$. $J_x$ ($J_y$)
is the nearest neighbor hopping along the $x$ ($y$) direction.
With a uniform magnetic field $B$ along the $z$ direction,
the flux per plaquette, in units of
the flux quantum $\Phi_0$, is $\phi=-Ba^2/\Phi_0$. 
Field $B$ gives
rise to the Peierls phase factor $e^{i2\pi n\phi}$ in the hopping. \\

Within  the Landau gauge, the above Hamiltonian has been engineered in cold atom experiments\cite{Ian}.
 In this case, the vector potential is given by
$A_x=0$ and $A_y = -\phi x$ resulting in a  Hamiltonian  that is cyclic in $y$. Therefore, 
the eigenstates of the system
can be written as $\Psi_{n,m}= e^{ik_y m} \psi_n $ where $\psi_n$ satisfies the Harper equation\cite{Hof}
\begin{equation}
e^{ik_x}\psi^r_{n+1}+e^{-ik_x} \psi^r_{n-1} + 2\lambda \cos ( 2 \pi n \phi+ k_y)\psi^r_n = E \psi^r_n .
\label{harper}
\end{equation}
Here $n$ ($m$) is the site index along the $x$ ($y$) direction, $\lambda=J_y/J_x$
and $\psi^r_{n+q} =\psi^r_n$, $r=1, 2, ...q$ are linearly independent solutions.
In this gauge, the magnetic Brillouin zone
is $ -\pi/qa \le k_x \le \pi/qa$ and $-\pi \le k_y \le \pi$.

At the rational flux $\phi=p/q$, where $p$ and $q$ are relatively prime integers, the energy spectrum has $q-1$ gaps.
These spectral gaps are labeled by two
quantum numbers which we denote as $\sigma$ and $\tau$. The integer $\sigma$ is the Chern number 
, the quantum number associated with Hall conductivity\cite{TKKN}  and 
$\tau$ is an integer.
For a Fermi level inside each energy gap, the system
is in an integer quantum Hall state\cite{QHE} characterized by its Chern number
$\sigma$ which gives transverse conductivity\cite{TKKN} $C_{xy}=\sigma \frac{e^2}{h}$.\\

The quantum numbers $(\sigma, \tau)$ satisfy a Diophantine equation (DE)\cite{Dana}, that applies to all 2DEG systems that exhibits magnetic translational symmetry,\\
\begin{equation}
\rho= \phi \sigma +\tau
\label{Dio}
\end{equation}
where $\rho$ is the particle density when Fermi level is in the gap.
For a given $\rho$ and $\phi$, there are infinity of possible solutions for where $[\sigma, \tau]$ are integers, given by,

\begin{equation}
[\sigma, \tau ] = [\sigma_0-n q, \tau_0 +  np ]
\label{DEsol}
\end{equation}

Here $\sigma_0, \tau_0$ are any two integers that satisfy the Eq. (\ref{Dio}) and $n$ is an integer.
The quantum numbers $\sigma$ that determine the quantized Hall conductivity correspond to
the change in density of states when the magnetic flux quanta in the system 
is increased by one and, whereas the quantum number $\tau$ is the change in density of states
when the period of the potential is changed so that there is one more unit cell in the system\cite{fractal1}.\\

For any value of the magnetic flux , the system described by the Hamiltonian (\ref{qh}), supports  only $n=0$ solution  of Eq. (\ref{DEsol}) for the
quantum numbers $\sigma$ and $\tau$. This is
due to the absence of any gap closing that is essential for topological phase transition
to states with higher values of $\sigma, \tau$. However, the DE which relates continuously varying quantities $\rho$ and $\phi$
with integers  $\sigma$ and $\tau$, has some important consequences about topological changes in close vicinity of rational values of $\phi$. We now show that the infinity of solutions
depicted in Eq.(\ref{DEsol}) reside in close proximity to the flux $\phi$  and label the fine structure of the butterfly. We illustrate this later in the paper by time-dependent perturbation that drives the Hamiltonian (\ref{qh})  periodically.\\


We begin by solving the Diophantine equation, using a geometrical approach well known in quasicrystal literature -- commonly referred as the
  ``{\it Cut and Projection Method}"\cite{Levine,QC}.
Note that although the explicit solution has been known\cite{Wiegmann}, our approach however  illustrates simplicity underlying the number theoretical
approach to solve this equation.
The basic idea is to obtain solutions by going to higher dimensions and the required solutions are the projections from two to one dimension.\\

We  start by defining two vectors:  a ``flux vector'' $\bm{F}$  and a ``topology vector $T_r$ as,

\begin{equation}
\bm{F}=(p,q),\,\,\quad
\bm{T}_r=[\sigma_r,\tau_r]
\end{equation}
 
 The DE is then rewritten as, 

\begin{equation}
r=\bm{F} \cdot \bm{T}_r
\label{diophantine}
\end{equation}

This implies that the gap index  $r$ is a  projection of the topology  vector onto the flux vector. However, this  projection
is  an integer.   This suggests the following scheme to obtain the components of $T_r$, namely $\sigma_r, \tau_r$ in terms of $r$ as follows.\\

\begin{figure}
\includegraphics[scale=0.3]{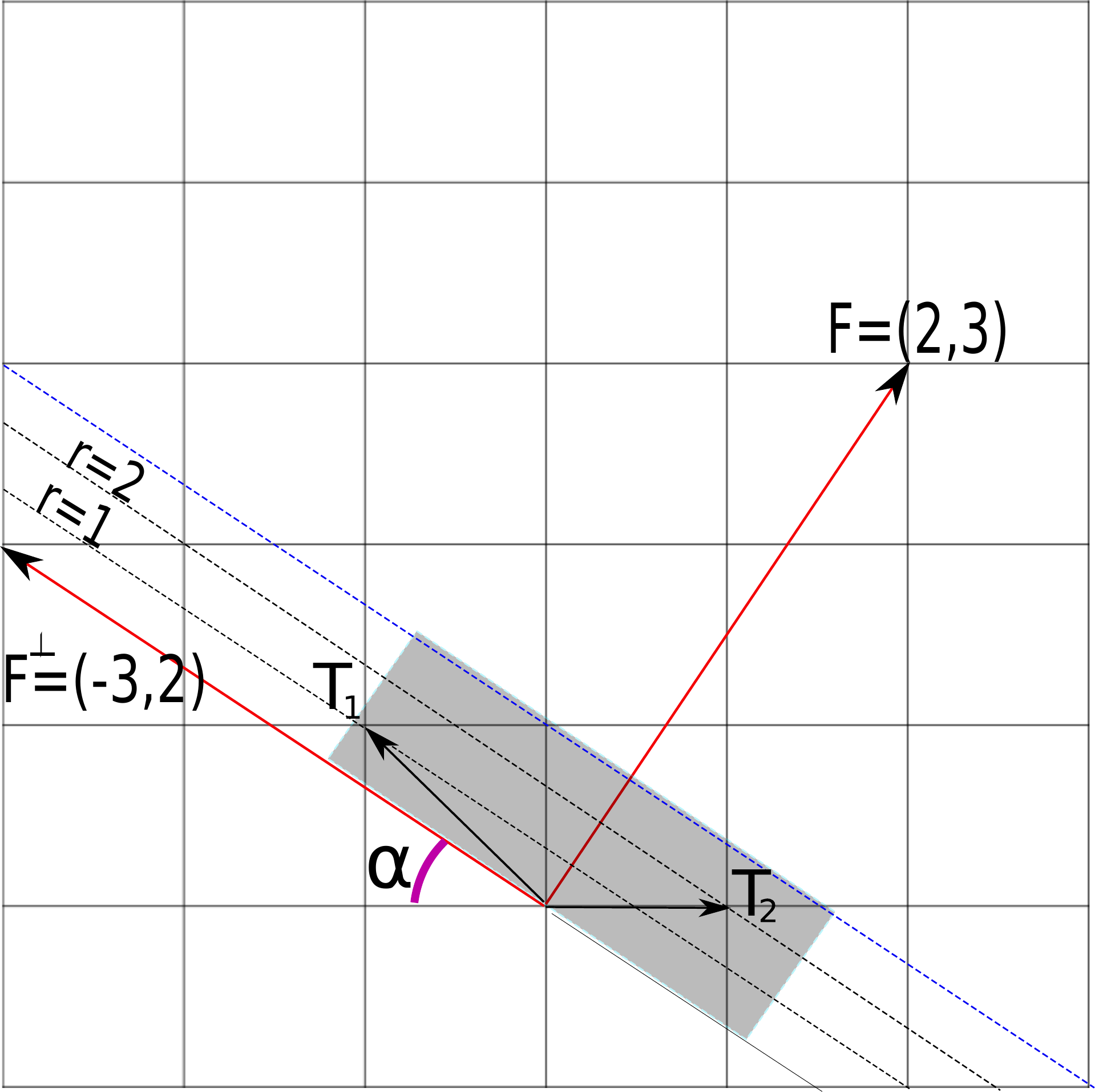}
\leavevmode \caption {Geometry associated with the cut and projection method to obtain the solutions of the Diophantine equation. In this example, the flux vector is chosen to give $\phi=2/3$. The set of parallel lines gives the solutions for each gap $r$. Two solutions are shown here, $\bf{T}_1$ and $\bf{T}_2$.}
\label{cut}
\end{figure}

As shown in Fig.  (\ref{cut}), we consider a two dimensional 
space with coordinates $(x,y)$.  A  square
lattice is defined  in this space by considering $x$ and $y$ at integer values. 

A vector  $\bm{F}$ is draw which points to square lattice point $(p,q)$. Then all possible solutions
of the Diophantine lay in the 2D square lattice, and are contained in a family of parallel lines
given by,

\begin{equation}
r=(p,q)\cdot(x,y)
\end{equation}

as indicated in Fig. \ref{cut}. To find integer solutions, we look at the family of parallel lines. These
lines are all perpendicular to the vector,

\begin{equation}
\bm{F}^\perp=(-q,p)
\end{equation}

which defines the line $y=-\phi x$, with slope $\phi$,i.e., $\tan \alpha= \phi$ where $\alpha$ is the angle between the $x$ axis and $\bm{F}^\perp$. If $x$ is chosen to be an integer, that we associate with a
Chern number, this will produce a $y$ coordinate,

\begin{equation}
y=-\phi \sigma_r
\end{equation}

However, although $y$ is in the family of parallel lines, it does not produce an integer.
But we find an integer just by taking the floor function of the previous equation,

\begin{equation}
\lfloor y \rfloor=-\lfloor \phi \sigma_r \rfloor
\end{equation}

so for each $\sigma_r$, the corresponding  $\tau_r$ is given by,

\begin{equation}
 \tau_r=-\lfloor \phi \sigma_r \rfloor
\end{equation}

Thus,  gaps are labeled by the coordinates of a two dimensional 
lattice,
\begin{equation}
 [\sigma_r,\tau_r]=[\sigma_r,-\lfloor \phi \sigma_r \rfloor]
\end{equation}

Furthermore, by using the identity $x=\lfloor x \rfloor+\left\{  x\right\} $ to express $\tau_r$ 
and inserting the solution into Eq. (\ref{diophantine}),
we obtain that,

\begin{equation}
 r=q\left\{ \phi \sigma_r \right\} 
 \label{map}
\end{equation}

Notice that care must be taken for negative Chern numbers, since $\left\{  -x\right\}=1-\left\{x\right\}$ for 
$x>0$.  We now define the {\it Hull function} $f$ as,
\begin{equation}
f(\phi, \sigma )= \{\phi \sigma_r\}
\label{hull}
\end{equation}
which is the filling factor for a Chern number at a given $\phi$.

This  formula can be inverted using the same methodology giving the Chern numbers as a function of the gap index,

\begin{equation}
 \sigma_r=\frac{q}{2}-q\left\{ \phi r +\frac{1}{2}\right\} 
 \label{invmap}
\end{equation}

\begin{figure*}
\includegraphics[width =.7\linewidth,height=0.85\linewidth]{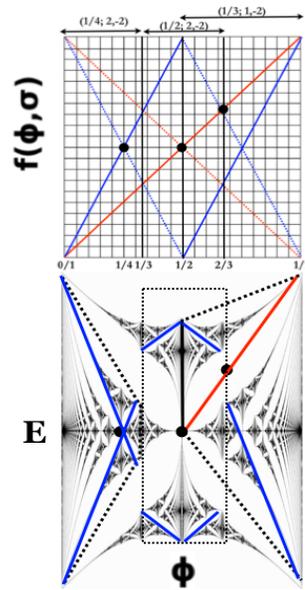}
\leavevmode \caption { The skeleton and the butterfly graphs showing explicitly the Cherns $\pm 1$ (red) and $\pm 2$ (blue) as a function of $\phi$. Positive Cherns are solid lines and negative Cherns are broken lines. 
In addition to the entire butterfly, we identify three butterflies in the graph, whose centers are shown with black circles and are marked in the bottom graph with a trapezoid or a rectangle. The bottom butterfly graph
shows these butterflies enclosed  inside the trapezoids:  $(2,-2)$ centered at $\phi=1/4$  and ($1,-2$) centered at $1/3$. In addition, there is another butterfly
enclosed in a rectangle, centered at $1/2$. The  flux values for three different butterflies are also shown with double arrowed lines that are labeled by the filling fraction and the Cherns of the butterfly in the upper graph.}
\label{HB}
\end{figure*}

The hull function\cite{Naumis1}  can be viewed as a kind of ``skeleton butterfly" plot that  encodes the  topological structure of the Hofstadter spectrum as we explain below. 
Earlier studies have discussed this skeleton in terms of the integrated density of states\cite{CW}. In this paper, we use this Hull function along with the numerically obtained  butterfly to lay out the topological patching of the entire butterfly.
Upper and lower graphs in Fig. \ref{HB}  illustrate the relationship between the butterfly graph and its skeleton version obtained from the Hull function.
We emphasize that although the quantitative analysis of the actual energies requires a numerical exploration, many features can be obtained using the Hull function. As we discuss below, this includes not only the dominant gaps but the fine structure associated with them.
Fig. \ref{HB} shows the filling fractions $r/q=f(\phi,\sigma)$   as a function 
of flux $\phi$ for Chern numbers $\pm 1, \pm 2$.  
Notice that each $f(\phi,\sigma)$  is 
just a saw-tooth function $\sigma$ distinct branches. The intersection of two branches are points where distinct Cherns meet. \\


Let us consider two successive gaps in the butterfly landscape, one with gap index $r=a$ , Chern $\sigma_a$ and another with gap index $r=b$ , Chern $\sigma_b$ emanating from the left of the  graph  and
meeting at a certain $\phi=p/q$. These two gaps will exchange their corresponding value of $r$ at the meeting point. From the Diophantine equation with $\phi \rightarrow \phi+ \delta \phi $, we obtain,
\begin{equation}
 \lim_{\delta \phi  \rightarrow 0^{-}}(\left\{ (\phi-\delta \phi )\sigma_a \right\}= \lim_{\delta \phi \rightarrow 0^{-}}\left\{ (\phi-\delta \phi)\sigma_b \right\})
\label{leftlimit}
 \end{equation}

This equation can only be satisfied provided,
\begin{equation}
 \sigma_a=\sigma_b+q
 \label{meeting}
\end{equation}
since the fractional part function has period $1$ and $\left\{x\right\}=x$ for $0\le x<1$,i.e., when applied to this particular case,
$\left\{\phi \sigma \right\}=\left\{(p/q) \sigma \right\}=(p/q) \sigma$ and the period is $q$. Thus the arguments of the fractional parts 
in Eq. (\ref{leftlimit}) can differ up to multiples of the period.

This equation, which we will refer as the  Chern meeting formula relates the topological quantum numbers of two swaths of the butterfly that meet at a point.

We now identify a ``central butterfly" and its fine structure as follows (see Fig. (\ref{HB})).  The central butterfly is the meeting of the two  smallest Cherns ( in magnitude ) and the value of $\phi$  where they meet  forms the center of that butterfly.
However, there is a whole  set of other Cherns that converge at such point. Such Cherns will be identified with the fine structure of this central butterfly as we will explain later.\\

Having identified the center, which we label as $p_c/q_c$, we now ask where are the boundaries of this butterfly.  A close inspection of the skeleton graph shows that
the boundaries of the central butterfly are the the closest intersections  labeled by the  largest Chern, larger than the  Chern that labels the butterfly. This is illustrated in Fig. \ref{HB}.

 \begin{figure}[htbp]
\includegraphics[width = .75\linewidth,height=.5\linewidth]{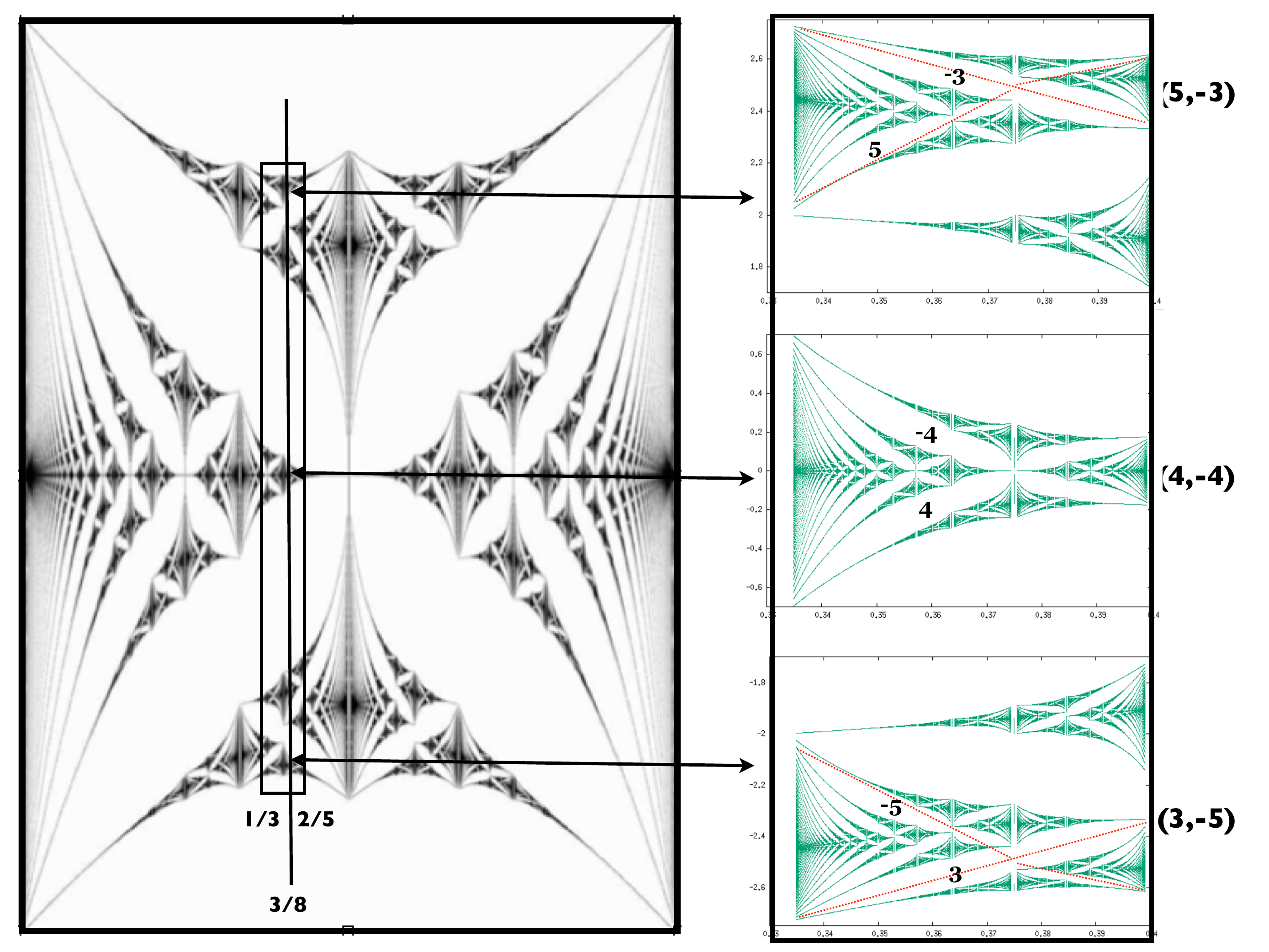}
\leavevmode \caption{ Three butterflies, all centered at $\phi=3/8$  with edges at $1/3$ and $2/5$ . The blowup of these butterflies ( in green)
 labeled with their Cherns is shown on the right. For off-centered butterflies,
thin red lines aid the eyes to identify the butterflies. We note that the topological states  of the three butterflies are linked by $\sigma_+-\sigma_- = q_c =8$. }
\label{LRchernC}
\end{figure}

 \begin{figure}[htbp]
\includegraphics[width = .75\linewidth,height=.5\linewidth]{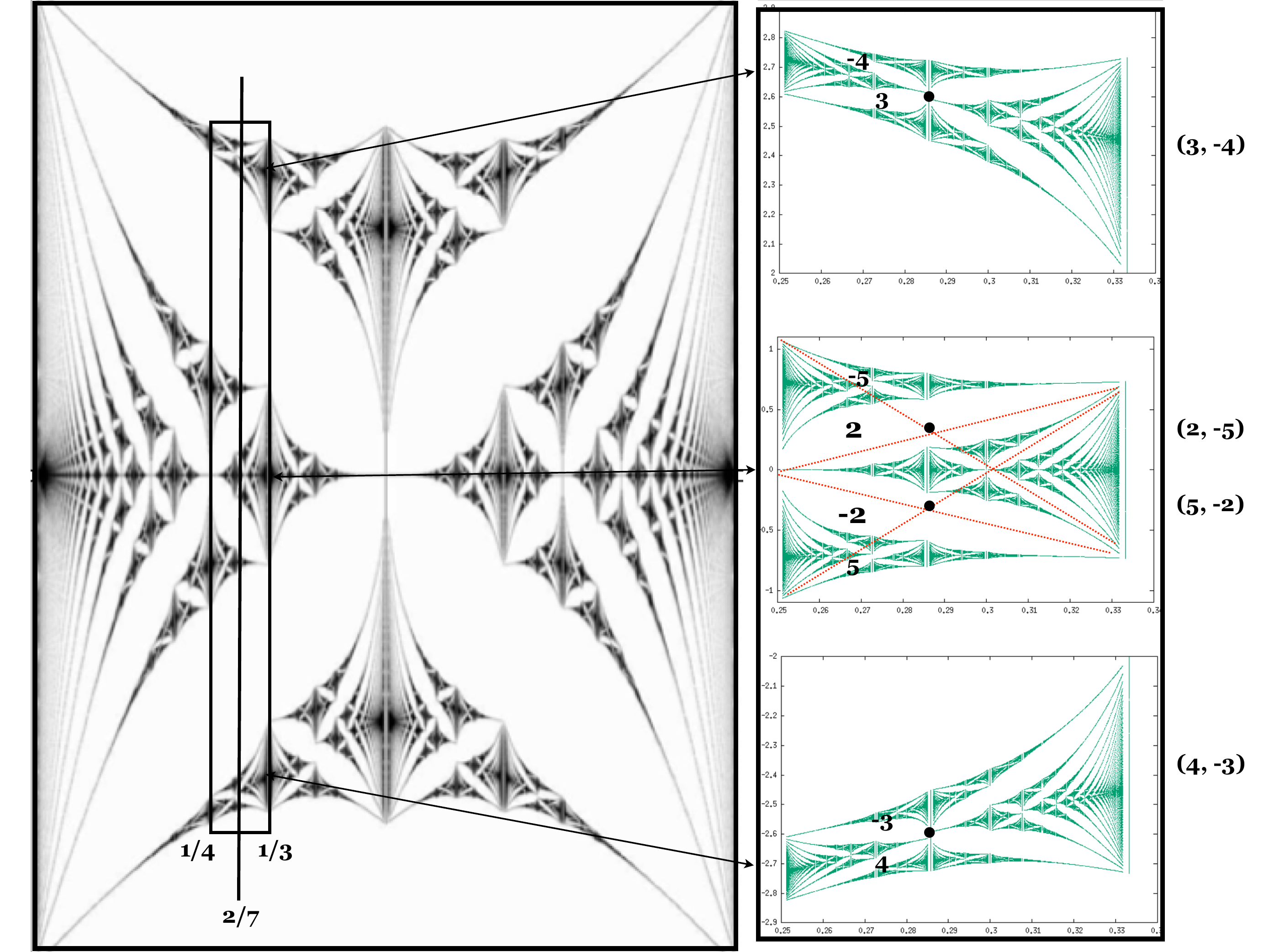}
\leavevmode \caption{ Four butterflies, all centered at $\phi=2/7$  with edges at $1/4$ and $1/3$. The blowup of these butterflies ( in green)
 labeled with their Cherns is shown on the right. For off-centered butterflies near $E=0$,
thin red lines with black dot as the center aid the eyes to identify the butterflies. We note that the topological states  of the four butterflies are linked by $\sigma_+-\sigma_- = q_c =7$. }
\label{LRchernO}
\end{figure}

\begin{figure*}
\includegraphics[width =.95\linewidth,height=0.5\linewidth]{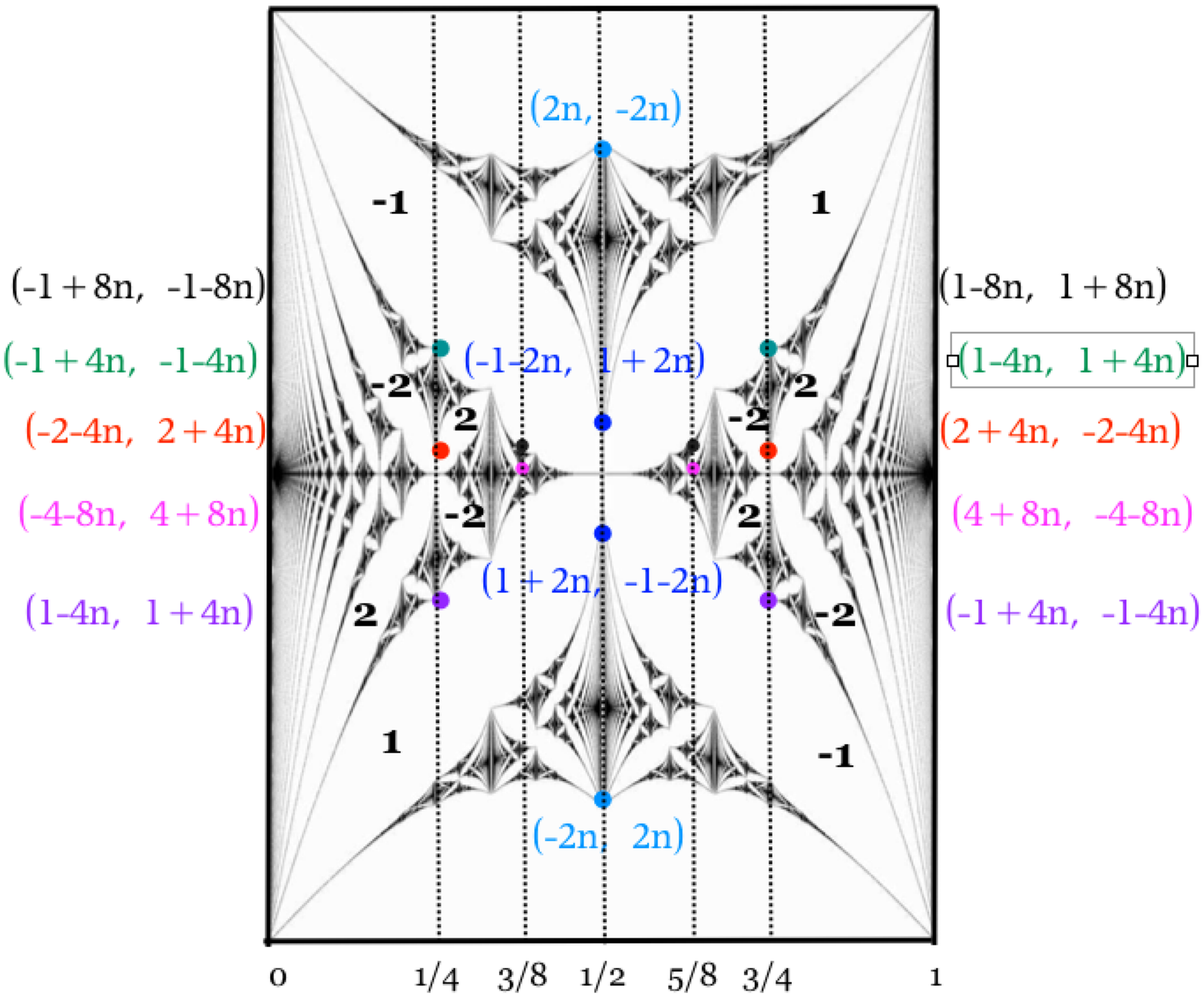}
\leavevmode \caption {Labeling some of the gaps of the  butterfly plot with their Chern numbers. Figure illustrates that the higher $n$-solutions of the Diophantine equation for a given flux  reside in close vicinity to that flux.  
Corresponding to certain rational flux values, shown explicitly on the horizontal axis, the color coded dots represent gaps whose left and right Cherns are showed inside the parentheses, labeled with matching colors.}
\label{fine}
\end{figure*}

The precise rules for locating the boundaries of the butterfly specified by a triplet of rationals labeling the left boundary, the center and the right boundary $(p_L/q_L, p_c/q_c, p_R/q_R)$ 
 were discussed in an earlier paper\cite{indu} which we state as follows.
 For the central  butterfly,  the Farey neighbors of $p_c/q_c$ are $p_L/q_L$ ( left neighbor)  and $p_R/q_R$ ( right neighbor)
   where $ q_L < q_c$ and $q_R < q_c$. These left and right neighbors of $p_c/q_c$ define the left and the right boundaries of the butterfly.  In other words,

\begin{equation}
\frac{p_c}{q_c} = \frac{p_L+p_R}{q_L+q_R}   \equiv  \frac{p_L}{q_L} \bigoplus \frac{p_R}{q_R}
\label{FT}
\end{equation}

We note that Farey neighbors satisfy the so called ``{\it friendly number}" equation,

\begin{equation}
p_c q_x - p_x q_c = \pm 1
\end{equation}
where $x = L, R$.

We now address the question of  determining the Chern numbers of every butterfly in the butterfly digram,  obtaining  the topological map of the entire butterfly graph. Every butterfly 
in this graph is characterized by a pair of Chern numbers,
which we will denote as $(\sigma_{+}, \sigma_{-})$, representing the Chern numbers of the two diagonal wings of the butterfly.

The simplest case happens to be the butterflies that are centered on the horizontal axis of the butterfly graph where the centers of such butterflies are described by flux values $p_c/q_c$ where $q_c$ is an
even integer. As shown in earlier paper\cite{indu}, and also follows from
the formula (\ref{map}),  such butterflies are characterized by  a pair Chern numbers  $(\sigma_{+}, \sigma_{-} =  (\frac{q_c}{2}, -\frac{q_c}{2})$.\\
 Another simple case of topological structure are the  butterflies located at the minimum or the maximum values of energy in the butterfly diagram. It is easy to show that these butterflies are characterized by Chern pairs
 $(\pm q_R,  \mp q_L)$.\\

In general, as shown in Fig. \ref{LRchernC} and Fig. \ref{LRchernO}, with every Farey triplet $(p_L/q_L, p_c/q_c, p_R/q_R)$ that satisfy the Eq. \ref{FT},
there are $q_x$ butterflies where $q_x = Min(q_L, q_R)$. These butterflies share the same value of the magnetic flux $\phi$ for their center and the boundaries, but are displaced in energy. The Chern number of this
family of butterflies are related by the solutions of the Chern meeting formula ( Eq. \ref{meeting} ). In other words, these butterflies are topologically linked by the equation,
\begin{equation}
|\sigma_{+}-\sigma_{-}| = q_c
\end{equation}

Fig. \ref{LRchernC} and Fig. \ref{LRchernO} provide examples of possible Chern numbers of such butterflies centered at $3/8$ and $2/7$.
 Below, we discuss a  corollary of the Diophantine equation that facilitates the precise
rules for determining these solutions and laying out the detailed topological map of the butterfly at all values in $E$ and $\phi$.


 To determine the topological map of the hierarchical set of gaps,  we  study the fine structure of the butterfly landscape determining the Chern numbers near a flux $\phi_0=p_0/q_0$. 
To do this, we do a simple ``titling" of the flux and $\rho$.  We substitute in the DE $\phi = \phi_0+\delta \phi$ and $ \rho = \rho_0 + \delta \rho$, and the
corresponding quantum numbers as $\sigma = \sigma_0+\Delta \sigma$ and $\tau = \tau_0 +\Delta \tau$. 
Now, taking the limits as $\delta \phi$ and $\delta \rho$ go to zero, we obtain,

\begin{equation}
\phi_0 \Delta \sigma+ \Delta \tau = 0;\,\,\,\,
\frac{\Delta \sigma}{\Delta \tau}= -\frac{q_0}{p_0}
\label{Dchern}
\end{equation}

We will refer this equation as the ``corollary " of the DE equation.\\

Since both $\Delta \sigma$ and $\Delta \tau$ are integers and $p_0$ and $q_0$ are relatively prime, the simplest solutions of Eq. (\ref{Dchern})
are $\Delta \sigma = \pm n q_0$ and $\Delta \tau = \mp n p_0$, where $n=0,1,2,...$. These solutions describe
the fine structure of the butterfly near a flux $\phi_0$. 

The above  corollary determines  the entire topological map of the butterfly as described in
Fig. (\ref{fine}). In addition to the dominant gaps, this plot  illustrates the fine structure both near the center as well as at the boundaries of every central butterfly in the entire butterfly landscape. As seen in the Fig. near $\phi =1/2$,
we see two cascades of gaps , characterized by upper and lower set of Cherns, forming a kind of fountain with fountainhead located at $E=0$ and $E_{max}$. \\


\begin{figure*}
\includegraphics[scale=0.3]{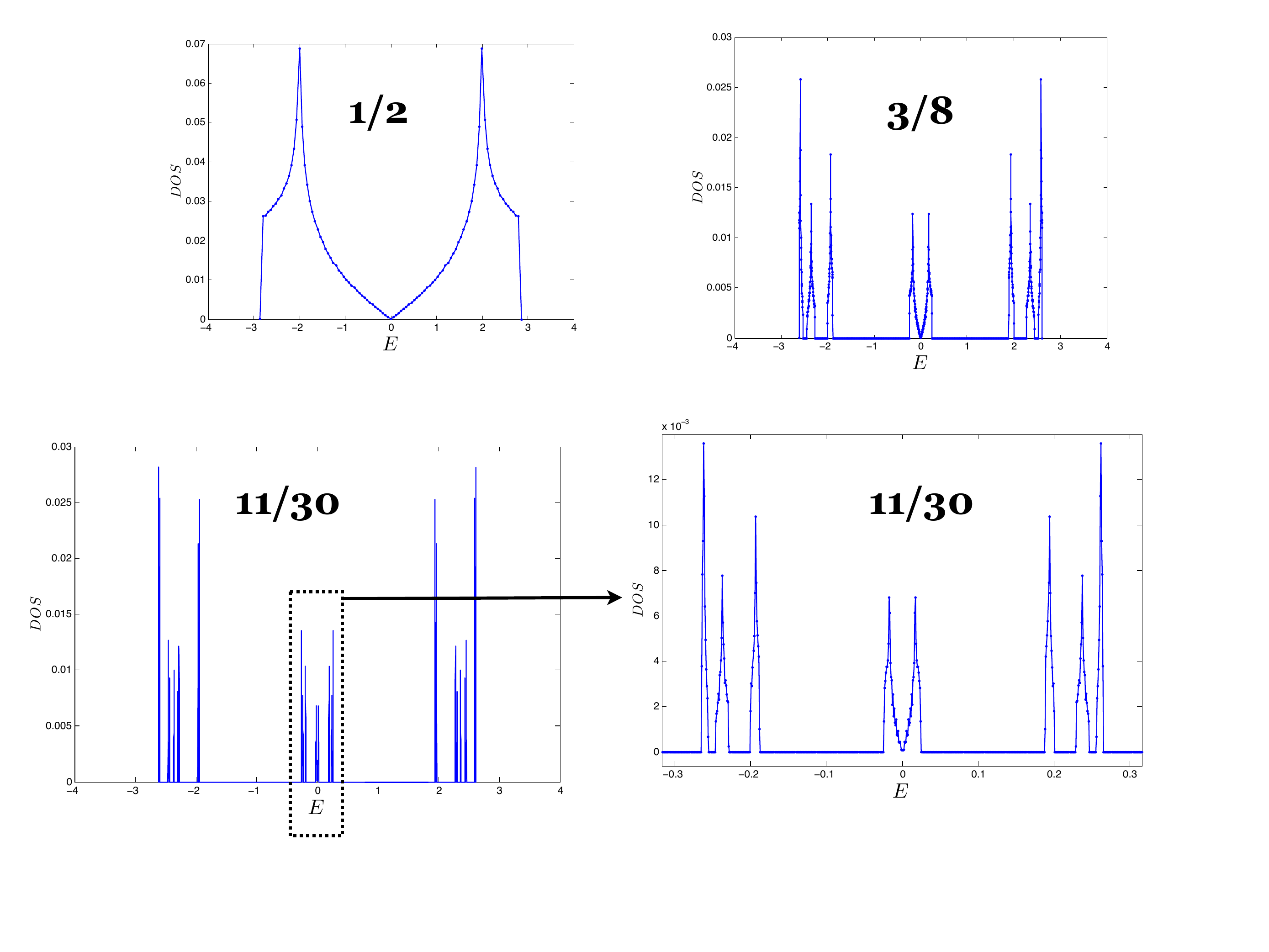}\\
\includegraphics[scale=0.3]{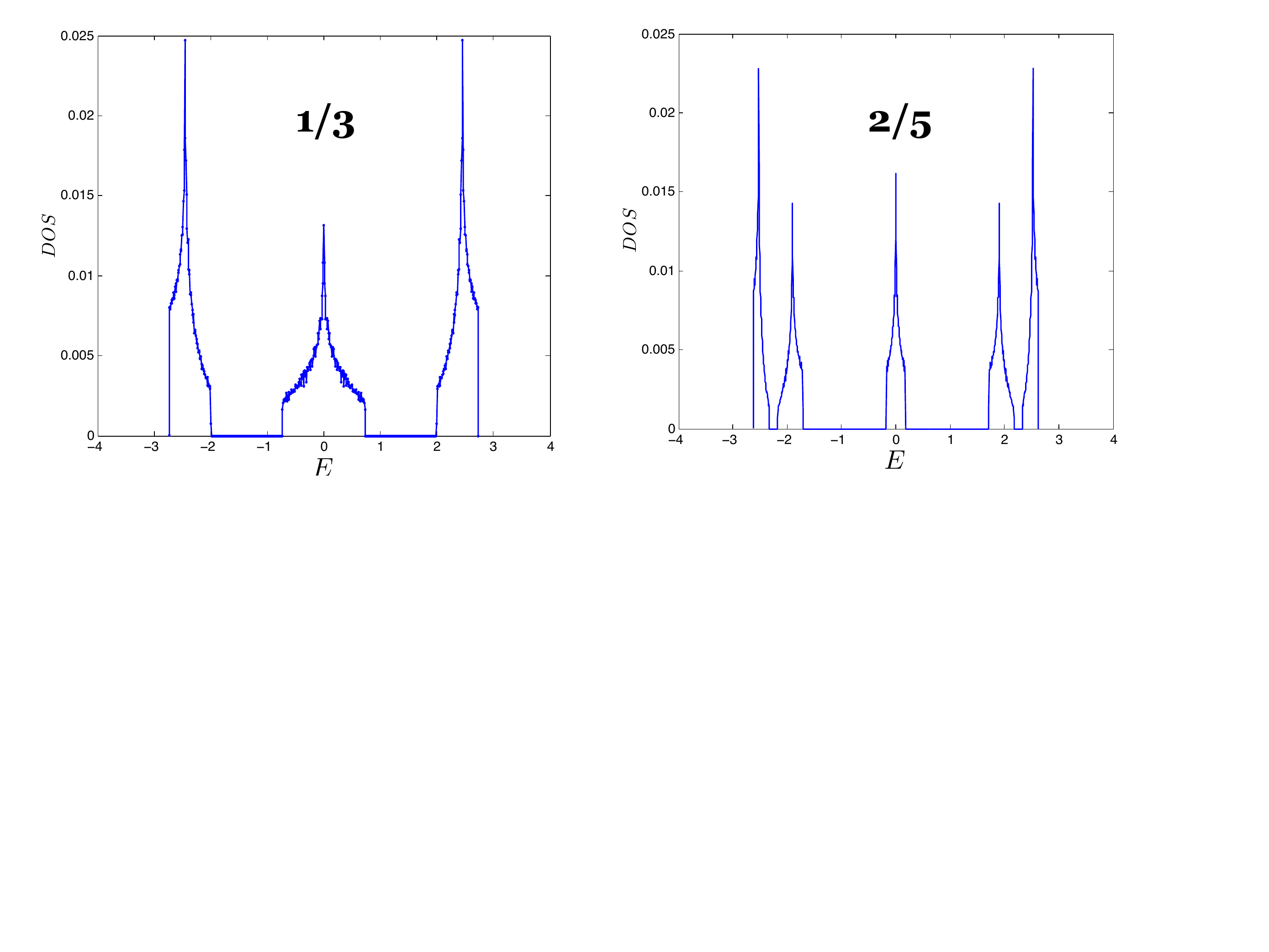}
\leavevmode \caption {(color on line)  Density of states (DOS) as a function of the energy for several fluxes.  This plot illustrates how Van-Hove singularities exist at every band center, irrespective of 
its location in flux value and energy.  Furthermore, they form self-similar patterns as 
 for the butterfly centered at $3/8$ and with left and right boundaries at $1/3$ and $2/5$. The next butterfly in this hierarchy\cite{indu} has its center at $11/30$. For odd $q$, the central band is a scaled version of the square lattice,
 as shown here for $\phi=1/3$ and $\phi=1/5$
}\label{VHevenodd}
\end{figure*}

\begin{figure*}
\includegraphics[scale=0.4]{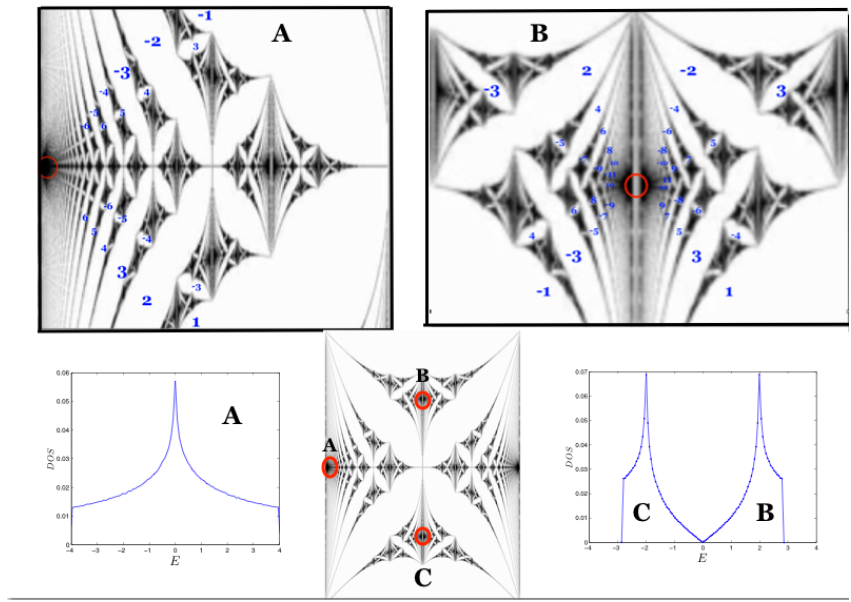}
\leavevmode \caption {(color on line) Illustrating the topological collapse near a Van-Hove singularity. Circles (red) show Van Hove singularities where a sequence of cascades of gaps with both positive and negative Cherns
annihilate.}
\label{VH}
\end{figure*}

\begin{figure*}
\includegraphics[scale=0.4]{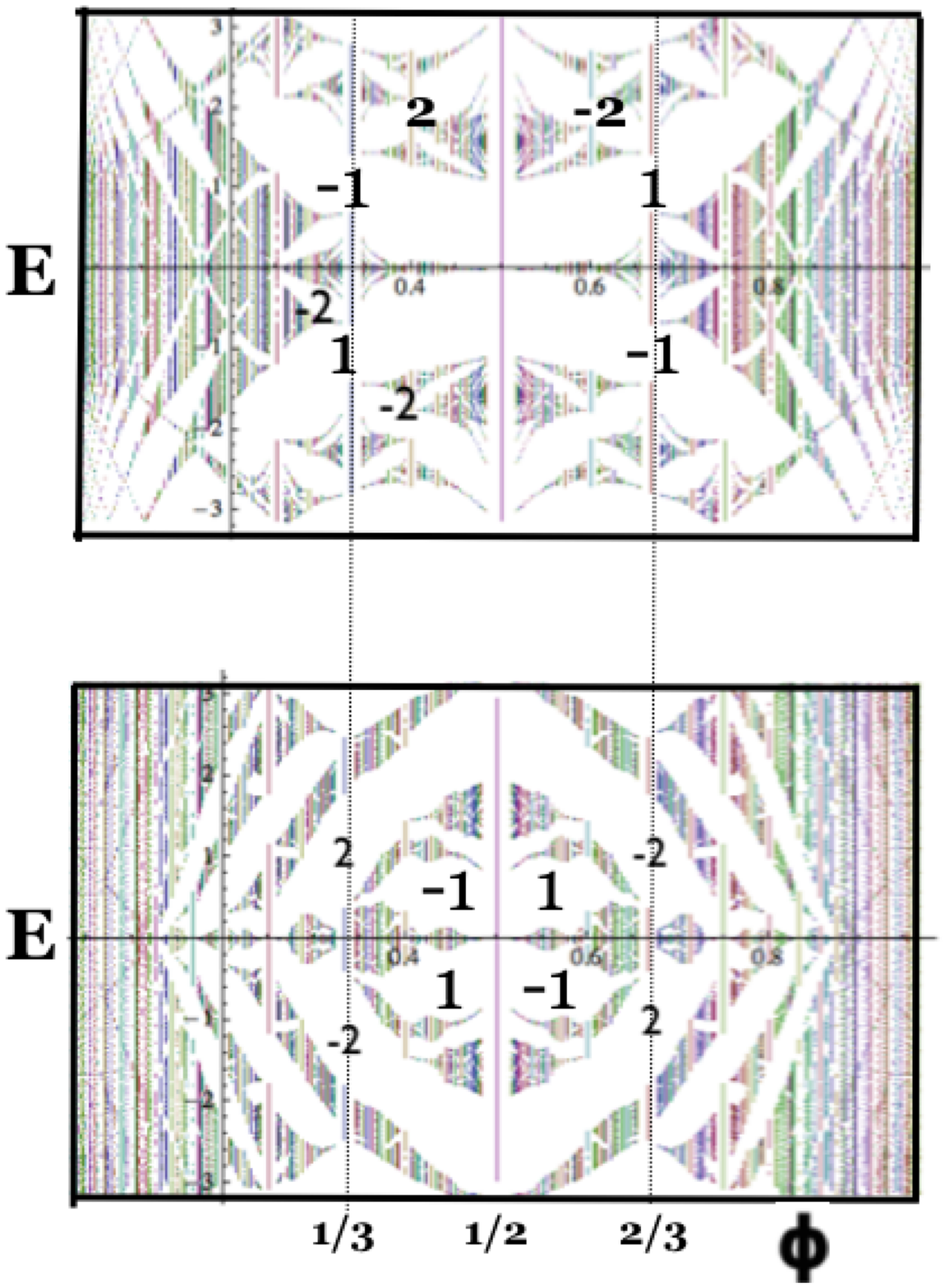}
\leavevmode \caption {(color on line) By varying the hopping along $y$-direction periodically, we can introduce a phase transition where the gap at $p/q=1/3$ and its vicinity is transformed from
a Chern-$\pm 1$ to the Chern $\mp 2$ state.  In other words $1 \rightarrow 1-3$ at $1/3$,  consistent with the corollary II as $\pm 1 = 1 \mp 3$.}
\label{EK}
\end{figure*}


Having determined the topological structure of the butterfly landscape where the gaps have been the  focus of our discussion, we now turn our attention to some interesting characteristics of the bands. We  show that
the fine structure of the gaps discussed earlier are rooted in Van Hove singularities that reside at the band center.
In the presence of a magnetic flux $p/q$,  a single band is split into $q$-bands.  The center of each of this sub-band  exhibits
a Van Hove singularity. Therefore, accompanying the hierarchical set of bands in the butterfly landscape are also a set of Van Hove singularities, seen  just with an eyeblink, as dark spots  -- that is,  high density points.
In fact low magnetic flux limit, discussed in the beginning applies to the neighborhood of every rational flux. 

Fig. (\ref{VHevenodd}) shows the Van Hove singularities at some of band centers in the fragmented spectrum in the presence of magnetic field.  This clearly implies that Van Hove are integral part of {\it every} band center,
irrespective of its location in energy. In other words,
no matter how fragmented a band is, its center is always a critical point that hosts a Van Hove singularity.  This is a consequence of the generic existence of saddle points in periodic two dimensional systems \cite{Jones}.  Consequently, in the case of incommensurate flux where bands have zero measure as band width of every band
approaches zero, the  surviving  Cantor set or ``dust" encodes  a fractal set of Van Hove singularities.  The figure shows zoomed versions of the DOS of the butterfly in the interval $1/3-2/5$. The nested set of plots
provide an estimate of energy scaling,
which turns out to be close to $10$, in agreement with an earlier estimate.\cite{indu}\\

We next address the question of how the topology of the butterfly  is influenced by the Van Hove singularities.
Fig. (\ref{VH}) shows the topological landscape in the vicinity of Van Hove singularities, illustrating what happens to
the zero-field Van Hove as the system is subjected to a small magnetic flux $\phi$. The magnetic field that fragments the band and the  resulting cascade of channels or gaps are characterized by
positive and negative Cherns interlaced as illustrated in the Fig. Again, this type of behavior is present at all band centers and Fig. (\ref{VH})  also shows a new Van Hove singularity and the topological landscape in its neighborhood at $\phi=1/2$.\\

Note that the sequence of Cherns near the Van Hove are the higher order solutions of the Diophantine equation, described in Eq. (\ref{Dchern}). This brings us to an interesting point about the importance of these
topological states that are crawling around the Van Hove.  It turns out that under  perturbations of the system,  these solutions take over, replacing the low Chern states with higher Cherns
which then become the dominant gaps in the system. \\

We illustrate this 
by perturbing such a systems, to induce quantum phase transitions to topological states
with $n > 0$ given by (\ref{DEsol}) with dominant gaps characterized by higher Chern numbers. We study butterfly spectrum for a
periodically kicked quantum Hall system\cite{ML}
where $J_y$ is a periodic function of time $t$ with period-$T$,
$J_y= \lambda \sum_n \delta(t/T-n)$, a system that was recently investigated\cite{ML}. Readers are referred to the original paper for various details of the system.\\

Figure shows a topological phase transition from Chern $\pm 1$ state to Chern $\mp 2$ state, seen clearly in simple flux values such as $1/3$ and $2/3$. The new topological state corresponds to 
the $n=1$ solution of the DE as shown in Eq.  (\ref{DEsol}).  This can be interpreted as a ``gap amplification"  as the Chern-$2$ gap after the phase transition has clearly become a dominant gap in the butterfly fractal
of the driven system.  This process where tiny gaps of the Hofstadter butterfly can be amplified may provide a possible pathway to see fractal aspects of
the  butterfly and engineer states with large Chern numbers experimentally. \\ 

In summary, the  topological characterization of the butterfly landscape is encoded in simple rules  that determine the topological map of the butterfly  at all energy and flux scales.  The entire hierarchy of
miniature butterflies that exists at any 
scale is accompanied by an orderly set of cascades of topological states that  attribute topological character to the Van Hove singularities.
Near these singularities,  higher -$n$ solutions of  the Diophantine equation play a central role. These solutions become dominant  under perturbations that
will induce topological phase transitions to higher Chern states in the system.\\

G. Naumis would like to thanks a PASPA-DGAPA UNAM sabbatical scholarship to spend a semester at George Mason and
 he acknowledges George Mason University for the hospitality. This work was partially 
funded by UNAM DGAPA-PAPIIT proyect 102513.

\end{document}